\documentstyle[12pt]{article}
\newcommand\beq{\begin{eqnarray}}
\newcommand\eeq{\end{eqnarray}}

\renewcommand{\theequation}{\thesection.\arabic{equation}}
\newcommand\la{\langle}
\newcommand\ra{\rangle}

\def\kone{k_{1}}
\def\ktwo{k_{2}}
\def\vep{\varepsilon}
\def\konehat{\hat{k}_{1}}
\def\ktwohat{\hat{k}_{2}}
\def\phat{\hat{p}}
\def\dpol{\frac{1}{2\vep^{2}}}
\def\spol{\frac{1}{2\vep}}

\def\slash#1{\rlap/{#1}}
\def\koneslash{\slash{\mkern-1mu \kone}}
\def\ktwoslash{\slash{\mkern-1mu \ktwo}}
\def\pslash{\rlap/{\mkern-1mu p}}
\def\pslash{\slash{\mkern-1mu p}}
\def\kslash{\slash{\mkern-1mu k}}
\def\delslash{\slash{\mkern-1mu \Delta}}
\def\Dk{\frac{d^{D}\kone}{(2\pi)^{D}}
\frac{d^{D}\ktwo}{(2\pi)^{D}}}

\begin{document}
\setlength{\baselineskip}{0.2in}
\begin{titlepage}
\noindent 

\noindent
\vspace{36pt}

\vspace{5pt}
\Large

\centerline{\bf Next-to-leading order $Q^2$-evolution}
\centerline{\bf of the transversity distribution $h_1(x, Q^2)$}

\normalsize

\par             
\par\bigskip
\par\bigskip
\vspace{0.7cm}                                                 

\centerline{A. Hayashigaki, Y. Kanazawa and Yuji Koike}
\centerline{\it  Graduate School of Science and Technology,
Niigata University, Ikarashi, Niigata 950-21, Japan}
\vspace{1in}

\centerline{\bf Abstract}
We present a calculation of the two-loop anomalous dimension
for the transversity distribution $h_1(x,Q^2)$, 
$\gamma^{h(1)}_n$, in 
the MS scheme of the dimensional regularization.  
Due to the chiral-odd nature, $h_1$ does not mix with
the gluon distributions, and thus our result is
the same for the flavor-singlet and nonsinglet distributions.
At small $n$ (moment of $h_1$),
$\gamma^{h(1)}_n$ is 
significantly larger than $\gamma^{f(1)}_n$ (the
anomalous dimension for the nonsinglet $f_1$), 
but approaches $\gamma^{f(1)}_n$
very quickly at large $n$, keeping the relation
$\gamma^{h(1)}_n > \gamma^{f(1)}_n$.
This feature is in parallel to the relation between
the one-loop anomalous dimension for $f_1$ and $h_1$.

\vspace{1.0cm}


\end{titlepage}

\newpage
\section{Introduction}
\setcounter{equation}{0}
\renewcommand{\theequation}{\arabic{section}.\arabic{equation}}
\hspace*{\parindent}

It is well known that the nucleon (spin-1/2 hadron) has three
independent twist-2 quark distributions $f_1(x, Q^2)$,
$g_1(x, Q^2)$ and $h_1(x, Q^2)$.  These distributions have a simple 
parton model interpretation and appear as a leading contribution
to various hard (semi-) inclusive processes.
The first two distributions, $f_1$ and $g_1$, are chiral-even and 
can be 
measured 
through the totally inclusive deep inelastic lepton-nucleon 
scattering (DIS).  In fact, accumulated DIS data 
on $f_1$ and $g_1$
has provided us with a rich information about the parton 
distributions
in the nucleon.
The third one, $h_1$, is chiral-odd and cannot be
measured through the totally inclusive DIS.  
However, it appears
as a leading contribution to
the transversely polarized nucleon-nucleon Drell-Yan 
process\,\cite{RS,AM,JJ,CPR},
semi-inclusive 
productions of baryons and pion in the DIS\,\cite{AM,JJ},
{\it etc}.  
This spin-dependent quark distribution, named as
transversity distribution in \cite{JJ}, 
measures the probability
in a transversely polarized nucleon
to find a quark polarized parallel to the
nucleon spin 
minus the probability to find it oppositely polarized.
It thus 
supplies an 
information on the nucleon's spin structure not provided by
the $g_1$ distribution, and is
expected to be measured 
by RHIC in the near future.

The $Q^{2}$-evolution of these parton distribution functions
is calculable in perturbative QCD.
It is governed by the anomalous dimensions for the 
corresponding
twist-2 operators and the $\beta$-function.
Leading order (LO) $Q^2$-evolution
of $f_1$ and $g_1$
has been known for a long time\,\cite{GW} and 
the LO $Q^2$-evolution for $h_1$ was also studied in \cite{AM,KT}.
For the next-to-leading order (NLO) $Q^2$ evolution, we need 
two-loop anomalous dimensions.  So far two-loop 
anomalous dimension
was calculated for $f_1$\,\cite{FRS,FRS2,GLY,FLK,CFP,FP} and for
$g_1$\,\cite{MN,Vogel}.  They have been used
for the phenomenological parametrization of $f_1$\,\cite{f1} 
and $g_1$\,\cite{g1} to analyze accumulated experimental data.

The purpose of this paper is to 
calculate the two-loop anomalous dimension for 
$h_1$, following the
method of \cite{FRS,FRS2}.
We calculate the two-loop correction to the 
two-point Green function which imbeds the 
twist-2 operators for 
$h_1(x,Q^2)$.  The calculation is done in the MS scheme
in the dimensional regularization.

The construction of this paper is the following:
In section 2, we briefly recall 
the twist-2 operators for the distribution
$h_1$ and summarize the formal aspect of the next-to-leading 
order (NLO) 
effect in
the $Q^2$-evolution of $h_1$.
In section 3, we describe the actual procedure 
of our calculation.  
Although the method is in parallel with that of \cite{FRS, FRS2},
presence of the free Lorentz index $\mu$ in the
twist-2 operators for $h_1$ brings additional complication.
We shall describe this peculiar feature for $h_1$ in 
great detail, but
omit the details which are common to both $h_1$ and $f_1$.
In section 4, we present the results and discuss
characteristic features of the NLO $Q^2$-evolution of $h_1(x,Q^2)$.
Readers not interested in the technical details can directly
proceed to this section.
Section 5 is devoted to a brief summary.
In Appendix A, we present the
necessary integral formulas to express
the anomalous dimension contribution from each diagram.
In Appendix B, we give the contribution to the anomalous dimension
from each two-loop diagram.

\section{Twist-two operator for $h_{1}(x,Q^{2})$ and its 
renormalization}
\renewcommand{\theequation}{\arabic{section}.\arabic{equation}}
\setcounter{equation}{0}

The transversity distribution $h_1(x,Q^2)$ 
of the nucleon is defined by the following
relation\,\cite{JJ}:
\beq
& &\int{ d\lambda \over 2\pi}e^{i\lambda x} \la PS | \bar{\psi}(0)
\sigma_{\mu\nu}i\gamma_5 \psi(\lambda n) |_Q |PS\ra
= 2 \left[ h_1(x,Q^2)(S_{\perp\mu}p_\nu - S_{\perp\nu}p_\mu)/M 
\right.\nonumber\\
& &\left.\qquad\qquad + h_L(x,Q^2)M(p_\mu n_\nu - p_\nu n_\mu )
(S\cdot n) + h_3(x,Q^2)M (S_{\perp\mu}n_\nu - S_{\perp\nu}n_\mu) 
\right],
\label{eq2.1}
\eeq
where $|PS\ra$ is the nucleon (mass $M$) state specified by 
the four momentum
$P$ and the spin vector $S$, and two light-like vectors
$p$ and $n$ are introduced by the relation,
$P^\mu= p^\mu + {M^2 \over 2}n^\mu$ and $pn=1$.
$S^\mu$ is decomposed as $S^\mu= (S\cdot n)p^\mu + (S\cdot p)n^\mu
+ S_\perp^\mu$.  In (\ref{eq2.1}), the gauge link operator
to ensure the gauge invariance is suppressed.
$h_L$ and $h_3$ are twist-3 and -4 distributions which are not
in our interest here.
Taylor expanding the both sides of (\ref{eq2.1}) with respect to 
$\lambda$,
one can relates the $n$-th moments of $h_{1,L}$ to
the following local operator:
\beq
\theta_{\mu\nu\mu_{1}\ldots\mu_{n}} = {\cal S}_{n}\overline{\psi}
i\gamma_{5}\sigma_{\mu\nu}iD_{\mu_{1}}\ldots iD_{\mu_{n}}\psi,
\label{eq2.2}
\eeq
where ${\cal S}_n$ denotes the symmetrization among $\nu,
\mu_1,\cdots,\mu_n$,
and the covariant derivative $D_\mu = \partial_\mu -igT^a A^a_\mu$
restores the gauge invariance.
In particular, the $n$-th moment of $h_1$
is related to a tower of the twist-2 operators:
\beq
& &\int_{-1}^1\,dx\,x^n h_1(x,Q^2) = a_n(Q^2)
\label{eq2.3}\\
& &\langle PS | 
\overline{\theta}_{\mu\nu\mu_1\cdots\mu_n}(Q)\Delta^\nu
\Delta^{\mu_1}\cdots \Delta^{\mu_n}
| PS \rangle
\nonumber\\ 
& &\qquad\qquad
=\frac{2}{M} a_n(Q) \left(S_{\mu}\hat{P}^{n+1}-P_{\mu}
\hat{S}\hat{P}^{n}
      +\frac{2}{n+2}M^{2}\Delta_{\mu}
         \hat{S}\hat{P}^{n-1}\right),
\label{eq2.4}
\end{eqnarray}
where $\overline{\theta}_{\mu\nu\mu_1\cdots\mu_n}$ is defined 
as the traceless part of
$\theta_{\mu\nu\mu_1\cdots\mu_n}$ defined by the condition,
\begin{eqnarray}
g^{\mu\mu_{i}}\overline{\theta}_{\mu\nu\mu_{1}\ldots\mu_{n}}
=g^{\nu\mu_{i}}\overline{\theta}_{\mu\nu\mu_{1}\ldots\mu_{n}}
=g^{\mu_{i}\mu_{j}}\overline{\theta}_{\mu\nu\mu_{1}\ldots\mu_{n}}=0,
\label{eq2.5}
\end{eqnarray}
and we introduced a null vector $\Delta^\mu$ ($\Delta^2=0$) 
which kills off the trace terms in 
$\overline{\theta}_{\mu\nu\mu_1\cdots\mu_n}$ as usual.

For simplicity we symbolically use the notation $O_n$ for
$\overline{\theta}_{\mu\nu\mu_{1}\cdots\mu_{n}}$ in what follows
in this section.
The bare-($O_n^B$) and the renormalized-
($O_n(\mu)$) composite operators are related by the renormalization 
constant $Z_{n}(\mu)$ for $O_n$ as
\begin{eqnarray}
O_{n}(\mu)&=&Z_n^{-1}(\mu)O_{n}^{B}.
\label{eq2.6}
\end{eqnarray}
The scale dependence of $O_n(\mu)$ is obtained by solving the
renormalization group equation
\begin{eqnarray}
\mu\frac{dO_n(\mu)}{d\mu} + \gamma_n(g(\mu))
O_n(\mu)= 0,
\label{eq2.7}
\end{eqnarray}
where $\gamma_n(g(\mu))$ is the anomalous dimension
for the operator $O_n(\mu)$ defined as
\begin{eqnarray}
\gamma_n(g(\mu)) = \mu\frac{\partial}{\partial \mu}{\rm ln}
Z_n(\mu). 
\label{eq2.8}
\end{eqnarray}
This equation is solved to give
\begin{eqnarray}
O_n(Q^{2}) = O_n(\mu^{2}){\rm exp}
\left[-\int_{g(\mu^{2})}
^{g(Q^{2})}dg
\frac{\gamma_n(g)}{\beta(g)}\right]. 
\label{eq2.9}
\end{eqnarray}
Up to next-to-leading order, the anomalous dimension
$\gamma_n(\mu)$ and the beta function $\beta(g)$ can be expanded as
\begin{eqnarray}
\gamma_n(g)&=&{g^{2}\over 16\pi^2}\gamma_n^{(0)}
+{g^{4}\over (16\pi^2)^2}\gamma_n^{(1)}+O(g^{6}), 
\label{eq2.10}\\
\beta(g)&=&- {g^{3}\over 16\pi^2} \beta_0
-{g^{5}\over (16\pi^2)^2} \beta_1 
+O(g^{7}),
\label{eq2.11}
\end{eqnarray}
where the coefficients of the $\beta$-functions
are well known;  $\beta_0=11- (2/3)N_f$, $\beta_1=102-(38/3)N_f$,
with the number of quark flavor $N_f$.
Inserting these equations into (\ref{eq2.9}), one obtains
the next-to-leading order evolution equation for the $n$-th moment
of $h_1$ as
\begin{eqnarray}
{{\cal M}_n[h_{1}(Q^{2})]\over{\cal M}_n[h_{1}(\mu^{2})]}=
\left(\frac{g^{2}(Q^{2})}
{g^{2}(\mu^{2})}\right)^{\frac{\gamma^{(0)}_n}{2\beta_{0}}}
\left(\frac{\beta_{0}+\beta_{1}g^{2}(Q^{2})/16\pi^2}
{\beta_{0}+\beta_{1}g^{2}(\mu^{2})/16\pi^2}\right)
^{\frac{1}{2}\left(\frac{\gamma_{n}^{(1)}}
{\beta_{1}}-\frac{\gamma_{n}^{(0)}}{\beta_{0}}\right)}, 
\label{eq2.12}
\end{eqnarray}
with
\beq
{g^2(Q^2)\over 16\pi^2}={1 \over \beta_0{\rm ln}(Q^2/\Lambda^2)}
\left[ 1- { \beta_1 {\rm ln}{\rm ln}(Q^2/\Lambda^2)
\over \beta_0^2 {\rm ln}(Q^2/\Lambda^2) } +\cdots \right],
\label{eq2.12b}
\eeq
where we introduced the shorthand notation
${\cal M}_n[h_1(\mu)]\equiv\int_{-1}^1\,dx\,x^nh_1(x,\mu)$.
This NLO effect in $h_1$ has to be combined with
the NLO effect in the hard cross section in the parton level
to give a prediction for a relevant physical
cross section in the NLO level.  

In the MS scheme of the dimensional regularization,
one obtains $Z_n(\mu)$ as
\beq
Z_n(\mu) = 1 + \sum_{k=1}^\infty { X^n_k(g(\mu)) \over \varepsilon^k},
\label{eq2.13}
\eeq
with $\varepsilon = (4-d)/2$.
Then the anomalous dimension is obtained as
\beq
\gamma_n(g)= -2g^2 { \partial X^n_1 \over \partial g^2}.
\label{eq2.14}
\eeq
Therefore
$\gamma^{(1)}_n$ in (\ref{eq2.10}) 
is $-4$ times
the coefficient 
of the $g^4/(16\pi^2)^2$ term in $X^n_1(\mu)$.
(See \cite{FRS} for the detail.)

\section{Calculation of the two-loop anomalous dimension} 
\setcounter{equation}{0}
\renewcommand{\theequation}{\arabic{section}.\arabic{equation}}

To obtain the renormalization constants for 
$\bar{\theta}^{\mu\nu\mu_1\cdots
\mu_n}$ in (\ref{eq2.4}), we calculate the two-loop correction
to the two-point function for $\bar{\theta}$
with off-shell quark lines having the momentum
$p$ ($p^2 < 0$).  The calculation of the two-loop Feynman diagram
is rather involved, and we follow the procedure of \cite{FRS,FRS2}.
To discuss renormalization of $\bar{\theta}$, the presence of
$i\gamma_5$ is irrelevant, since 
$i\gamma_5\sigma^{\mu\nu} =-1/2 \epsilon^{\mu\nu\kappa\rho}
\sigma_{\kappa\rho}$.
We shall henceforth omit $i\gamma_5$ 
for notational simplicity and
consider the renormalization
of the operator $O_n^{\mu\nu\mu_1\cdots\mu_n}$ defined by
\beq
O_n^{\mu\nu\mu_1\cdots\mu_n}={\cal ST}_{n}\overline{\psi}
\sigma^{\mu\nu}iD^{\mu_{1}}\cdots iD^{\mu_{n}}\psi,
\label{eq3.1}
\eeq
where ${\cal ST}_n$ denotes the symmetrization among $\nu, 
\mu_1,\cdots
\mu_n$ and the traceless condition in (\ref{eq2.5}).
As usual, we contract
$O_n^{\mu\nu\mu_1\cdots\mu_n}$ with a null vector
$\Delta^\mu$ to kill off the trace terms, 
and consider the renormalization of 
$O_n^\mu\cdot\Delta\equiv O_n^{\mu\nu\mu_1\cdots\mu_n}
\Delta_\nu\Delta_{\mu_1}\cdots\Delta_{\mu_n}$ in the following.
The calculation is
done 
in the Feynman gauge for the gluon propagator, but the results
should be independent of the gauge.

We first need the Feynman rule for the basic 
vertices for the operator $O_n^\mu\cdot\Delta$ 
with $0$, $1$, $2$ gluon lines
shown in Fig.1:
\beq
O_n^{(0)\mu}\cdot\Delta &=& \sigma^{\mu\nu}\Delta_\nu\hat{p}^n,
\label{eq3.2}\\
O_n^{(1)\mu}\cdot\Delta &=& gT^a\sigma^{\mu\nu}
\Delta_\nu\sum_{j=1}^n
\hat{p'}^{j-1}\Delta^\lambda\hat{p}^{n-j},
\label{eq3.3}\\
O_n^{(2)\mu}\cdot\Delta&=& g^2 \sigma^{\mu\nu}\Delta_\nu
\left[\sum_{1 \le j < l \le n} \hat{p'}^{j-1}
(\hat{p}+\hat{k'})^{l-j-1}
\hat{p}^{n-l}\Delta_\lambda\Delta_\rho T^aT^b\right.\nonumber\\
& &\qquad\qquad \left.
+ \sum_{1 \le j < l \le n} \hat{p'}^{j-1}(\hat{p}+\hat{k})^{l-j-1}
\hat{p}^{n-l}\Delta_\lambda\Delta_\rho T^bT^a\right],
\label{eq3.4}
\eeq
where 
we introduced the notation $\hat{p}\equiv p\cdot\Delta$
for a four vector $p$.
If one replaces $\sigma^{\mu\nu}\Delta_\nu$ by $\delslash$
in (\ref{eq3.2})-(\ref{eq3.4}), one obtains the Feynman rule
for the same basic vertices for $f_1$, which are used in \cite{FRS}.
Relevant two-loop 
diagrams are shown in Figs. 2, 3 and 4.  Diagrams shown in Fig. 3
are only for the flavor-singlet 
distribution, but they are identically zero
for $h_1$ owing to its chiral-odd
nature.  Those in Fig. 4 are the self energy corrections
and are given in \cite{FRS}.

Among the contribution from each diagram in Fig. 2, we can show
that the figs. (d), (g), (h), (l), (m), (q) and (r) give 
the same anomalous dimension to $h_1$ as the $f_1$ distribution
calculated in \cite{FRS}.
To see this, we note the relation
\beq
\sigma^{\mu\nu}\Delta_\nu = i\gamma^\mu\delslash -i\Delta^\mu
=-i\delslash \gamma^\mu +i\Delta^\mu.
\label{eq3.5}
\eeq
The term proportional to $\Delta^\mu$ is the trace term, 
and thus we can
drop them and use either $i\gamma^\mu\delslash$ or
$-i\delslash\gamma^\mu$ instead of $\sigma^{\mu\nu}\Delta_\nu$
in the actual calculation.  
We shall henceforth use the three forms for the
vertex
factor, $\sigma^{\mu\nu}\Delta_\nu$, $\gamma^\mu\delslash$, 
$-\delslash\gamma^\mu$, interchangeably (we also drop factor $i$
for simplicity) without mentioning explicitly.
Therefore the difference in the Feynman rule
between $h_1$ and $f_1$ is merely the presence of $i\gamma^\mu$
next to $\delslash$.
The loop corrections 
in the above diagrams are located in the right or left 
half of the diagram and hence the presence of $\gamma^\mu$ does
not bring any 
change in the calculation of the renormalization constants 
compared with the $f_1$ case.
We can thus take the results from \cite{FRS} for these diagrams.

It is also easy to confirm that the diagrams (a), (b) and (c)
become finite due to the form of the vertex, and hence no
contribution to the anomalous dimension
unlike the case for $f_1$.  Diagram (p) 
has no contribution as in the case of $f_1$.
Calculation of (e) and (f) is also standard.  One gets
contribution from these two diagrams which are independent of $n$
unlike $f_1$ case.

Rather heavy calculation 
is required for the diagrams
(i), (j), (k), (n) and (o).  Compared with the 
renormalization of $f_1$,
additional complication 
is caused 
by the presence 
of the free Lorentz index $\mu$ in $O_n^\mu\cdot \Delta$.
To explain our strategy 
of the calculation, we take the diagram Fig. 2 (i) as 
an example.  We assign the momentum as shown in Fig. 5.
Application of the Feynman rule results in 
the following contribution
from the diagram.
\beq
& &F^{(i)} = g^4{1 \over 2N_c} C_F \int Dk_1 \int Dk_2
{ \hat{k_1}^n N(k_1, k_2) \over D(k_1, k_2) }, 
\label{eq3.21}\\[10pt]
& &\qquad D(k_1,k_2) = k_1^4 k_2^2 (k_1-p)^2 (k_2-p)^2 (k_1- k_2)^2, 
\label{eq3.22}\\[10pt]
& &\qquad N(k_1,k_2) = \gamma_\lambda \koneslash 
\sigma^{\mu\nu}\Delta_\nu
\koneslash\gamma_\rho
(\koneslash - \ktwoslash ) \gamma^\lambda
(\pslash-\ktwoslash) \gamma^\rho,
\label{eq3.23}
\eeq
where 
\beq
\int Dk_1\equiv \int {d^dk_1 \over (2\pi)^d}.
\eeq
Our task is to 
extract from (\ref{eq3.21}) the $1/\varepsilon$ pole terms
proportional to 
$\sigma^{\mu\nu}\Delta_\nu \hat{p}^n$.
To this end we first contract the index $\rho$ in 
the numerator (\ref{eq3.23}),
noting the properties of the $\gamma$ matrix in $d=4-2\varepsilon$
dimension.
This gives 
\beq
N(k_1,k_2) &=& -2(1+\varepsilon)\gamma_\lambda\koneslash
\sigma^{\mu\nu}\Delta_\nu\koneslash (\pslash 
-\ktwoslash)\gamma^\lambda
(\koneslash-\ktwoslash)\nonumber\\[5pt]
& & +4\varepsilon (\koneslash-\ktwoslash)
\koneslash
\sigma^{\mu\nu}\Delta_\nu\koneslash (\pslash -\ktwoslash)
\nonumber\\[5pt]
& & +4\varepsilon (\pslash-\ktwoslash)
\koneslash
\sigma^{\mu\nu}\Delta_\nu\koneslash (\koneslash -\ktwoslash)
\nonumber\\[5pt]
& & -4\varepsilon (k_1-k_2)\cdot (p-k_2) \gamma_\lambda
\koneslash
\sigma^{\mu\nu}\Delta_\nu\koneslash\gamma^\lambda.
\label{eq3.24a}
\eeq
Noting $\gamma_\lambda
\koneslash
\sigma^{\mu\nu}\Delta_\nu\koneslash\gamma^\lambda =
-2\varepsilon
\koneslash
\sigma^{\mu\nu}\Delta_\nu\koneslash$, the last 
term of (\ref{eq3.24a}) 
becomes $O(\varepsilon^2)$ and cannot produce 
$1/\varepsilon$-poles
so that
we can drop it.
Using the same relation in the first term,
one gets up to $O(\varepsilon)$
\beq
N(k_1,k_2)&=& -4(\pslash-\ktwoslash)\koneslash
\sigma^{\mu\nu}\Delta_\nu \koneslash (\koneslash-\ktwoslash)
-4\varepsilon \koneslash
\sigma^{\mu\nu}\Delta_\nu \koneslash (\pslash-\ktwoslash)
(\koneslash-\ktwoslash)
\nonumber\\[5pt]
& &+4\varepsilon (\koneslash-\ktwoslash)\koneslash
\sigma^{\mu\nu}\Delta_\nu \koneslash (\pslash-\ktwoslash).
\label{eq3.24b}
\eeq
In the first and the third terms 
in (\ref{eq3.24b}), we can discard the
terms which have $\pslash$ at the leftmost or the rightmost
places, since
those terms are either ultraviolet finite from power counting or
only cause trace terms which are proportional to $\Delta^\mu$.
In fact, for example, one can write
\beq
& &\int Dk_1 \int Dk_2 { \hat{k_1}^n \pslash \koneslash \gamma^\mu
\delslash
\koneslash \ktwoslash \over D(k_1,k_2) }
= A \pslash \gamma^\mu \hat{p}^{n+1}(p^2)^{-(1+2\varepsilon)}
+B p^\mu \pslash \delslash \hat{p}^n (p^2)^{-(1+2\varepsilon)}
\nonumber\\[5pt]
& &\qquad\qquad +B' p^\mu \hat{p}^{n+1} (p^2)^{-(1+2\varepsilon)}
+C \Delta^\mu\pslash\delslash \hat{p}^{n-1}
(p^2)^{-2\varepsilon}
+C' \Delta^\mu \hat{p}^{n}
(p^2)^{-2\varepsilon},
\label{eq3.24}
\eeq
where $A$, $B$ and $B'$ terms are finite, and $C$ and
$C'$ terms are the trace terms.
Dropping these terms, one can rewrite (\ref{eq3.24b}) as follows:
\beq
N(k_1,k_2) &=& -4k_1^2 \ktwoslash \koneslash \delslash \gamma^\mu
-4\varepsilon k_1^2 \gamma^\mu \delslash \koneslash \ktwoslash
-4(1-\varepsilon) \ktwoslash \koneslash \gamma^\mu \delslash
\koneslash \ktwoslash
\nonumber\\[5pt]
& &+4\varepsilon ( 2k_1 \cdot k_2 -2p\cdot k_1+ 2 p\cdot k_2 -k_2^2)
\koneslash \gamma^\mu \delslash \koneslash
-4\varepsilon k_1^2 \koneslash \gamma^\mu\delslash \ktwoslash.
\label{eq3.26}
\eeq
The integration of the first and the second terms in (\ref{eq3.26})
can be handled as follows.  One can write 
\beq
\int Dk_1 \int Dk_2 { k_1^2 \hat{k}_1^n \ktwoslash \koneslash \delslash
\gamma^\mu \over D(k_1, k_2) }
= A\delslash\gamma^\mu + ({\rm finite\ as}\ \varepsilon \to 0),
\label{eq3.27}
\eeq
where $A$ can contain a pole at $\varepsilon = 0$.
The $A$ term can be projected out by multiplying $p^\mu$ to both sides of
(\ref{eq3.27}) and taking trace of both sides.
This is equivalent to replace in the numerator of (\ref{eq3.27}) as
\beq
\ktwoslash \koneslash \delslash \gamma^\mu
\rightarrow { 1 \over \hat{p} } ( \hat{p} k_1 \cdot k_2 
+ \hat{k}_1 p\cdot k_2 - \hat{k}_2 p\cdot k_1 )\delslash\gamma^\mu.
\label{eq3.28}
\eeq
Similarly to (\ref{eq3.28}), we can replace in the second term of
(\ref{eq3.26}) as
\beq
\gamma^\mu \delslash\koneslash \ktwoslash  
\rightarrow { 1 \over \hat{p} } ( \hat{p} k_1 \cdot k_2 
+ \hat{k}_1 p\cdot k_2 - \hat{k}_2 p\cdot k_1 )\gamma^\mu\delslash.
\label{eq3.28b}
\eeq
After the replacement (\ref{eq3.28}) 
in (\ref{eq3.27}), the integration can be easily
done by the same method as \cite{FRS}:
For example, $k_1\cdot k_2$ in the first term of (\ref{eq3.28})
is replaced by $1/2(k_1^2 + k_2^2 -(k_1 -k_2)^2)$.
Then each of these three terms combined with the $k_1^2$
factor in (\ref{eq3.27}) cancel some of the factors 
in the denominator $D(k_1, k_2)$ in (\ref{eq3.22}),
leaving the $O(k_{1,2}^8)$ factor in the denominator.
After this cancellation, one needs only one Feynman
parameter for each integration ($\int Dk_1$ and $\int Dk_2$)
and thus one  
can easily carry out the integration.

As is shown in this example,
our strategy of the calculation is
to devise a suitable replacement for each term in the numerator
to project out the pole terms proportional to
$\gamma^\mu\delslash$ or $\delslash\gamma^\mu$,
ignoring the trace term ($\sim \Delta^\mu$).
After this projection, it turns out that we
can express the final formula for the pole contributions
of the two-loop diagrams in terms of the integral formula
$I_1$ to $I_{14}$
listed in the appendix of \cite{FRS2} and two new integrals
$I_{15}$ and $I_{16}$ shown in the Appendix.

As was stated in (\ref{eq2.14}), simple pole terms ($\sim1/\varepsilon$)
from each two-loop Feynman diagram give rise to the anomalous dimension.
There are two 
origins for the simple poles: (i) $O(\varepsilon^0)$ contribution
in the numerator combined with the $1/\varepsilon$ singularity
from the integral. (ii) $O(\varepsilon)$ contribution
in the numerator combined with the $1/\varepsilon^2$ singularity
from the integral.  For the type-(i) singularity, one
can subtract the counter terms after transforming
each two-loop integral into the form of $I_1\sim I_{16}$.
Therefore, in 
the integral formulas given in \cite{FRS2}
and Appendix A, the counter terms for the simple pole terms are 
already subtracted.
To subtract the counter terms
for the type-(ii) $1/\varepsilon$-pole, we have 
to come back to the original diagram\,\cite{FRS}.
In Fig. 6,
we schematically showed the counter term
contribution for the diagram (i) in Fig. 2.
In Fig. 6(a), the counter-term corresponds to the renormalization
of the operator in one-loop (Fig. 6(b)), but the numerator
of the diagram in Fig. 6(a) has no $\varepsilon$.
In Fig. 6(c) the counter-term comes from the one-loop contribution
to the vertex correction (Fig. 6(d)). This time the numerator
of the diagram in Fig. 6(c) becomes $O(\varepsilon^2)$, hence no
counter-term contribution.  To summarize: we only have to
subtract the counter-terms for the simple poles in $I_1\sim I_{16}$
for the diagram (i).  This is also true for other diagrams
in Fig. 2.

The integration of the third term in (\ref{eq3.26}) is rather demanding.
We repeatedly apply the anti-commutation relation to move the 
leftmost $\ktwoslash$ untill it merges the righmost $\ktwoslash$
to form $\ktwoslash\ktwoslash =k_2^2$.
One then obtains
\beq
& &\ktwoslash \koneslash \gamma^\mu\delslash \koneslash \ktwoslash
= 2k_1\cdot k_2 \gamma^\mu\delslash\koneslash\ktwoslash
-4\hat{k}_1k_2^\mu\koneslash\ktwoslash +2k_1^2 k_2^\mu \delslash
\ktwoslash \nonumber\\[5pt]
& & \qquad\quad + 4 \hat{k}_2k_1^\mu\koneslash\ktwoslash
-2\hat{k}_2k_1^2\gamma^\mu\ktwoslash
-2k_1\cdot k_2\koneslash\gamma^\mu\delslash\ktwoslash
+k_2^2\koneslash\gamma^\mu\delslash\koneslash.
\label{eq3.29}
\eeq
The first term of (\ref{eq3.29}) can be handled in the same way as
(\ref{eq3.28b}).  

The fifth term of (\ref{eq3.29}) is the easiest to deal with.
One only has to make the replacement
\beq
\gamma^\mu\ktwoslash \rightarrow {p\cdot k_2 \over \hat{p}}\gamma^\mu
\delslash
\label{eq3.30}
\eeq
by the same argument as in (\ref{eq3.27}).

The third term of (\ref{eq3.29}) can be treated as follows.
To meet with more general situation we replace $k_2^\mu$ by $k_1^\mu$
($k_1^\mu$ can be put equal to 
$k_1^\mu$ or $k_2^\mu$ in the following formula.)
From power counting, one can write
\beq
\int Dk_1 \int Dk_2 { \hat{k}_1^n k_1^2 k_1^\mu \delslash
\ktwoslash \over D(k_1,k_2) } = A \delslash \gamma^\mu + B \Delta^\mu
\delslash\pslash + ({\rm finite\ as}\ \varepsilon \to 0),
\label{eq3.31}
\eeq
where $A$ and $B$ can contain poles at $\varepsilon=0$. 
To project out the $A$-term, we multiply 
$\pslash\gamma^\mu$ from the left and take the trace.  To $O(\varepsilon)$,
this is equivalent to the replacement,
\beq
k_1^\mu\delslash\ktwoslash \rightarrow {-1\over 2\hat{p}}
(1+\varepsilon)(\hat{k}_1p\cdot k_2 + \hat{k}_2 p\cdot k_1 -\hat{p}
k_1\cdot k_2)\delslash\gamma^\mu,
\label{eq3.32}
\eeq
in the numerator of (\ref{eq3.31}).
Similarly to (\ref{eq3.32}), we often meet
the following replacement in the calculation:
\beq
k_1^\mu\ktwoslash\delslash \rightarrow {-1\over 2\hat{p}}
(1+\varepsilon)(\hat{k}_1p\cdot k_2 + \hat{k}_2 p\cdot k_1 -\hat{p}
k_1\cdot
k_2)\gamma^\mu\delslash.
\label{eq3.32b}
\eeq

We can deal with the second and fourth term of (\ref{eq3.29}) as follows.
Without loss of generality one can write
($q^\mu = k_1^\mu$ or $k_2^\mu$)  
\beq
\int Dk_1 \int Dk_2 { \hat{k}_1^{n+1} q^\mu \koneslash \ktwoslash
\over D(k_1,k_2) } = A\gamma^\mu \delslash + B\Delta^\mu + 
C \Delta^\mu \pslash \delslash  +
({\rm finite\ as}\ 
\varepsilon \to 0),
\label{eq3.33}
\eeq
where $A$, $B$ and $C$ can contain poles at
$\varepsilon = 0$, and we can ignore finite contribution.   
To project out the $A$-term, we combine two operations.
(i) Multiply $p^\mu\delslash\pslash$ from the right
and take trace.   
(ii) Multiply $\gamma^\mu\pslash\hat{p}$ from the right
and take trace.  
Taking the difference between these two results, it turns out that
the projection of the $A$-term in (\ref{eq3.33}) is equivalent to
the following replacement in the integrand:
\beq
q^\mu\koneslash\ktwoslash \rightarrow {1 \over 2\hat{p}^2 }
(1+\varepsilon)\left\{
\hat{p}(p\cdot k_2 q\cdot k_1 - p\cdot k_1 q\cdot k_2)
-p\cdot q ( \hat{k}_1 p\cdot k_2 - \hat{k}_2 p\cdot k_1 ) 
\right\}\gamma^\mu
\delslash.
\label{eq3.34}
\eeq

The sixth term in (\ref{eq3.29}) can be rewritten as
$\koneslash\gamma^\mu\delslash\ktwoslash = 2k_1^\mu\delslash
\ktwoslash -2\hat{k}_1 \gamma^\mu\ktwoslash + \gamma^\mu\delslash
\koneslash\ktwoslash$ and we can apply (\ref{eq3.32}),
(\ref{eq3.30}), (\ref{eq3.28b}) for each term, and likewise
for the seventh term in (\ref{eq3.29}).

As an application of the above replacement 
rules, we prove a usefull
lemma here.

\vskip 0.3cm
\noindent
{\bf Lemma:} 
A term of the form $\varepsilon\kslash \sigma^{\mu\nu}\Delta_\nu
\kslash'$, where $k$ and $k'$ are either $k_1$ or $k_2$,
does not give rise $1/\varepsilon$ pole.

\vskip 0.2cm
\noindent
{\bf Proof:}
We write 
$\varepsilon\kslash \gamma^\mu\delslash
\kslash ' = 2\varepsilon k^\mu\delslash
\kslash ' -2\varepsilon\hat{k} \gamma^\mu\kslash' + 
\varepsilon\gamma^\mu\delslash
\kslash\kslash'$, and apply 
(\ref{eq3.32}),
(\ref{eq3.30}), (\ref{eq3.28b}) for each term.  Then it vanishes,
ignoring $O(\varepsilon^2)$ term.  This completes the proof.

\vskip 0.3cm
Applying this lemma, we can drop fourth and the fifth
terms in (\ref{eq3.26}), which completes transformation
of all terms in (\ref{eq3.26}) into the requested forms
as mentioned before.
By this lemma 
we can drop many of the terms in the calculation of
the diagrams (i), (j) and (k).

The obtained rules for the replacement 
(\ref{eq3.28}), (\ref{eq3.28b}), (\ref{eq3.30}),
(\ref{eq3.32}), (\ref{eq3.32b}), (\ref{eq3.34})
give all the necessary formula to carry out the integration
of the two-loop diagrams.
Using the method described in this section,
we obtained the anomalous dimension from 
all the diagrams in Fig. 2.
They are given in Appendix B.

\section{Results and discussions}
\setcounter{equation}{0}
\renewcommand{\theequation}{\arabic{section}.\arabic{equation}}

Collecting all the contribution from each diagram in Figs. 2 and 4
(See Appendix B and appendices of \cite{FRS, GLY}),
we obtained $\gamma^{(1)}_n$ for $h_1$ as,
\beq
\gamma^{h(1)}_n &=& 
4C_F^2 \left[ S_2(n+1)-2S_1(n+1) -{1\over 4}\right]\nonumber\\[5pt]
& &+C_FC_G\left[-16S_1(n+1)S_2(n+1) -
{58\over 3}S_2(n+1) + {572\over 9}S_1(n+1) -{20\over3}\right]
\nonumber\\[5pt]
& &-8\left(C_F^2 -{1\over 2}C_FC_G\right)
\left[4S_1(n+1)\left\{S_2'\left({n+1\over 2}\right)
-S_2(n+1) -{1\over 4}\right\}
-8\widetilde{S}(n+1)\right.\nonumber\\[5pt]
& &\left. + S_3'\left({n+1\over 2}\right)
-{5\over 2}S_2(n+1) + {(1+(-1)^n)\over (n+1)(n+2)} +{1\over 4}
\right]\nonumber\\[5pt]
& &
+{32\over 9}C_FT_R\left[3S_2(n+1)-5S_1(n+1) + {3\over 8}\right],
\label{eq4.0}
\eeq
where
\beq
S_k(n)&=&\sum_{j=1}^n{1\over j^k},\\[5pt]
S_k'\left({n\over 2}\right)&=&{1+(-1)^n \over 2}S_k\left({n\over 2}\right)
+{ 1-(-1)^n \over 2}S_k\left({n-1\over 2}\right),\\[5pt]
\widetilde{S}(n)&=&\sum_{j=1}^n{ (-1)^j \over j^2}S_1(j).
\eeq
(Note that $2/3 S_1(n)$ in the seccond line of (A.8) in \cite{GLY}
should read $2/3 S_3(n)$.)  This final formula for 
$h_1(x,Q^2)$ is very complicated 
as in the case of $f_1(x,Q^2)$ and is subject to numerical analysis
to study its $Q^2$-evolution.
In table 1, we present the actual numbers of $\gamma^{h(1)}_n$
($\gamma^{(1)}_n$ for $h_1$) 
for $n=0,...,20$.  This table can be 
compared with Table 2 of \cite{FRS}, which presents 
$\gamma^{f(1)}_n$ ($\gamma^{(1)}_n$ for $f_1$).  
(Note that $n$ in \cite{FRS} corresponds to our $n-1$.)
We also plotted $\gamma^{h,f(1)}_n$ for $N_f=3,\ 5$
in Fig. 7 to compare their behavior.  From Fig. 7, 
one sees clearly that at small $n$
$\gamma^{h(1)}_n$ is significantly larger than $\gamma^{f(1)}_n$
but approaches very quickly to $\gamma^{f(1)}_n$, keeping the condition
$\gamma^{h(1)}_n > \gamma^{f(1)}_n$.
This feature is the same as the one-loop anomalous dimensions
$\gamma^{h,f(0)}_n$ which reads
\beq
\gamma^{f(0)}_n &=& 2C_F \left( 1 - { 2 \over (n+1)(n+2)}
+ 4\sum_{j=2}^{n+1}{1 \over j}\right),
\label{eq4.1}\\
\gamma^{h(0)}_n &=& 2C_F \left( 1 
+ 4\sum_{j=2}^{n+1}{1 \over j}\right),
\label{eq4.2}
\eeq
and hence $\gamma^{h(0)}_n > \gamma^{f(0)}_n$ for all $n$.
Actually this tendency is even stronger for the two-loop case.
We note that for $n=0$ the anomalous dimension for $f_1$ is zero
in all orders because of the Ward identity for the vector
current.  On the other hand, $h_1$ projects onto the
tensor operator $\bar{\psi}\sigma^{\mu\nu}\psi$ for $n=0$,
for which there is no conservation law.  
Hence $\gamma^{h(0,1)}_0 \neq 0$.  This can be taken as a good 
reason for $\gamma^{h(0,1)}_n > \gamma^{f(0,1)}_n$
especially at small $n$.
From the difference between $\gamma^{f(1)}_n$ and $\gamma^{h(1)}_n$
at small $n$,
we expect that the NLO effect
leads to larger difference
in the $Q^2$-evolution in the small-$x$ region.

At large $n$, $\gamma^{h,f(1)}_n$ receives
dominant contributions from diagrams (d), (g), (h), (m), (q), (r)
in Fig. 2 which 
give the same anomalous dimension both for $h_1$ and $f_1$,
and sub-dominant 
contribution from (n), (l), (f), (e) and the self energy
corrections shown in Fig. 4.
All other diagrams in Fig. 2 gives minor corrections, although
they are very important at small $n$.
As is discussed in \cite{GLY}, $\gamma^{f(1)}_n$ behaves
as $\sim {\rm ln}(n)$ at large $n$ with the cancellation
among $\sim {\rm ln}^2n$ and $\sim {\rm ln}^3n$ terms which
arise from some of the diagrams.
Equation (\ref{eq4.0}) and
quick merging between $\gamma^{h(1)}_n$ and $\gamma^{f(1)}_n$
at large $n$ ( $n \geq 10$ )
with the relation $\gamma^{h(1)}_n > \gamma^{f(1)}_n$ for all $n$ 
clearly shows $\gamma^{h(1)}_n \sim {\rm ln}(n)$, which also
supports
the correctness of our calculation.

The relevant quantities for the $Q^2$-evolution 
of the moments are
$\gamma^{(0)}_n/2\beta_0$ 
and $\gamma^{(1)}_n/2\beta_1$ as is seen from
(\ref{eq2.12}).
In Fig. 8(a), 
we plotted these quantities for $f_1$ and $h_1$ with $N_f=3$.
Since the NLO effect is determined by their difference
(see (\ref{eq2.12})),
we plotted $\gamma^{f,h(1)}_n/2\beta_1-\gamma^{f,h(0)}_n/2\beta_0$ in
Fig. 8(b) for $N_f=3$ and $5$ cases. 
From Fig. 8(b), one expects that at small $n$ 
the NLO effect in the $Q^2$ evolution is quite different
between ${\cal M}_n[h_1(Q^2)]$ and ${\cal M}_n[f_1(Q^2)]$.
Since $\gamma^{f(0,1)}_n \rightarrow 0$ as $n\to 0$, 
$\gamma^{f(1)}_n/2\beta_1-\gamma^{f(0)}_n/2\beta_0$ abruptly drops to
zero as $n\to 0$.  But this is not the case for $h_1$:  The difference
between the black circles and squares in Fig. 8(a) shows
a characteristic behavior as shown in Fig. 8(b).

As an example of the $Q^2$-evolution,
we show in Fig. 9 (a) and (b) the $Q^2$-evolution 
of the tensor charge and 
the first moments, respectively, with the parameters $N_f=3$
and $\Lambda=0.232$ GeV in (\ref{eq2.12b}).
(Here we are interested in the NLO effect
in the anomalous dimension and the $\beta$-function, and thus
we adopted the same value for the $\Lambda$-parameter
in the LO and NLO evolution.)
They are normalized at $Q^2=1$ GeV$^2$.  
At $n=0$, only diagrams (e), (f), (i), (j), (k)
survive.  They give
the anomalous dimension for the tensor charge as
\beq
\gamma^{h(1)}_0 &=& -19C_F^2 +{257\over 9}C_F C_G - 
{52\over 9}C_F T_R\nonumber\\
&=& {724\over 9} -{104\over 27}N_f.
\label{eq4.3}
\eeq
From Fig. 9(a),
we can compare the LO and NLO
$Q^2$-evolution of the tensor charge of the nucleon.
One sees that the NLO effect is sizable 
as is expected from Fig. 8 (b).
In Fig. 9(b), we plotted the LO and NLO $Q^2$-evolution both
for ${\cal M}_1[f_1(Q^2)]$ and ${\cal M}_1[h_1(Q^2)]$.
Although the NLO effect in the anomalous dimension
(the second factor in 
the right hand side of (\ref{eq2.12})) makes this ratio 
smaller,
the NLO effect in the coupling constant (the first factor
in the right hand side of (\ref{eq2.12}))
completely cancels this
effect.  For $f_1$, the latter effect is actually 
bigger than the former effect.
As was stated in section 2, the NLO effect
in the distribution function has to be combined 
with the NLO effect in the hard cross section 
in the parton level to give
a prediction for a physical quantity.
We will pursue this issue in a future publication.

\vskip 1.0cm
\noindent
{\bf Table 1} The coefficients of the two-loop anomalous dimension 
$\gamma^{(1)}_n$ for $h_1$ with the number of quark flavors
$N_f=3,\ 4,\ 5$.
\begin{center}
\begin{tabular}{|c|r|r|r|} \hline
    \multicolumn{1}{|c}{\raisebox{-1.2ex}[0pt]{$n$}} 
& \multicolumn{3}{|c|}
           {$\gamma_{n}^{(1)}$} \\ 
   \cline{2-4}
      & \multicolumn{1}{c|}{$N_{f}=3$} 
      & \multicolumn{1}{c|}{4} 
      & \multicolumn{1}{c|}{5} \\ \hline
 
    0 &  68.89 &  65.04 &  61.19 \\
    1 & 100.00 &  92.00 &  84.00 \\
    2 & 123.08 & 111.92 & 100.76 \\
    3 & 140.51 & 126.83 & 113.15 \\
    4 & 155.36 & 139.60 & 123.83 \\
    5 & 167.87 & 150.33 & 132.79 \\
    6 & 179.02 & 159.93 & 140.84 \\
    7 & 188.84 & 168.38 & 147.92 \\
    8 & 197.81 & 176.12 & 154.43 \\
    9 & 205.91 & 183.11 & 160.31 \\
   10 & 213.43 & 189.61 & 165.79 \\
   11 & 220.34 & 195.58 & 170.82 \\
   12 & 226.82 & 201.19 & 175.56 \\
   13 & 232.84 & 206.40 & 179.96 \\
   14 & 238.54 & 211.34 & 184.14 \\
   15 & 243.88 & 215.97 & 188.06 \\
   16 & 248.96 & 220.38 & 191.79 \\
   17 & 253.77 & 224.55 & 195.32 \\
   18 & 258.36 & 228.53 & 198.71 \\
   19 & 262.72 & 232.32 & 201.92 \\
   20 & 266.91 & 235.96 & 205.01 \\ \hline
\end{tabular}
\end{center}

\section{Summary}
\setcounter{equation}{0}
\renewcommand{\theequation}{\arabic{section}.\arabic{equation}}

In this paper, we have carried out the calculation of
the two-loop anomalous dimension for the transversity 
distribution $h_1(x,Q^2)$.  The calculation was done in the 
MS scheme of the dimensional regularization.   
This completes the calculation of the anomalous dimensions
for all the twist-2 distributions of the nucleon in the NLO level.
We found $\gamma^{h(1)}_n$ is significantly larger 
than $\gamma^{f(1)}_n$
at small $n$, but approaches very quickly to $\gamma^{f(1)}_n$
at large $n$, keeping the condition
$\gamma^{h(1)}_n >\gamma^{f(1)}_n$.  This means
that the NLO effect for $h_1$ is quite different from that for $f_1$
in the small $x$-region.
As an example, we have 
compared the LO and the NLO $Q^2$-evolution of the tensor charge
and the first moment of $f_1$ and $h_1$.
We hope the peculiar feature of the $Q^2$-evolution
of $h_1$ studied in this paper will be measured in the 
ongoing and planned experiments in the near future.

\vskip 1.0cm
\noindent
{\bf Note added:}

After submission of this work for publication,
there appeared a preprint \cite{Vog} which calculated
the NLO splitting function for $h_1$ by a different method.  
Our whole result 
agrees with \cite{Vog}.
After submission, we also learned that there appeared
a preprint \cite{KM}, which calculated the two-loop
anomalous dimension for $h_1$ in the same formalism as ours,
but presented the result in a very complicated non-final form.
We further noticed that (1) their result for the diagrams Fig. 2 
(i), (j), (k) disagrees with ours, 
(2) their result for Fig. 2 (n) and (o) agrees
with our intermediate result but are not presented in a
simple form.  We believe something must be wrong
in their calculation.

\newpage

\appendix
\renewcommand{\theequation}{\Alph{section}.\arabic{equation}}
\renewcommand{\thesection}{\Alph{section}.}
\setcounter{equation}{0}
\centerline{\large{\bf APPENDIX}}

\section{Integral formula}
\setcounter{equation}{0}

To write down
the two-loop anomalous dimension
for $h_1$, it turns out that we need two more 
integral formula, $I_{15}$ and $I_{16}$, in addition to 
$I_1 \sim I_{14}$ defined in the appendix of \cite{FRS2}.
As in the case of $I_1 \sim I_{14}$, we define
them by subtracting the counter term contribution
for the simple-pole term but not subtracting the counter-term
for the double-pole term as explained in the text: 
\begin{eqnarray}
I_{15}(M,0) &\equiv& \int\Dk
\frac{\konehat^{M}\kone\cdot \ktwo}{\kone^{4}(\kone-p)^{2}
      (\ktwo-p)^{2}(\kone-\ktwo)^{2}} \nonumber \\
I_{15}(M,0)&=&  \frac{\phat^M}{(16\pi^{2})^{2}}
           \frac{3}{4(M+1)}\left[\;-\dpol
           + \spol\left\{\frac{1}{3M}+{1\over M+1}-2\right\}\right],
\quad(M\ge 1)\\[10pt]
I_{15}(0,0)&=&{-1 \over 4(16\pi^2)^2}\left( {1\over\varepsilon^2}
+{1\over\varepsilon}\right),\\[10pt]
I_{16}(M,N) &\equiv& \int\Dk
\frac{\konehat^{M}\ktwohat^{N}\kone\cdot p}{\kone^{4}(\ktwo-p)^{2}
(\kone-\ktwo)^{2}(\kone -\ktwo +p)^{2}}\nonumber \\
I_{16}(M,N)&=& \frac{\phat^{M+N}}{2(16\pi^{2})^{2}}\frac{M!N!}{(M+N)!}
          \sum_{k=0}^{N}\frac{(k+1)!(M+N-k-1)!}{(N-k)!(M+k+1)!}
          \left[\;-\dpol\right.\nonumber\\
      &&  +\left.\left.\left.\spol\right\{S_{1}(M+N-k-1)
           +S_{1}(M+k+1)-2S_{1}(M+N+1)\right\}\right],\nonumber\\
& &\qquad\qquad\qquad\qquad(M\ge 1,\ N\ge 0)\\
I_{16}(0,N)&=&{-\hat{p}^N \over (4\pi)^4}{N\over 4(N+1)}
\left[{1\over \varepsilon^2} 
+ {1\over \varepsilon}{1\over N}
\sum_{k=0}^{N-1}\left( 2S_1(N+1) \right.\right.\nonumber\\
& &\left.\left.-2S_1(k+1)+S_1(k)-S_{N-k-1}
-{k+1\over k+2}\right)\right],\qquad(N\ge 1)\\[10pt]
I_{16}(0,0) &=& 0,
\end{eqnarray}
where we discarded the finite contributions.  We also 
remind the readers
that our $\varepsilon$ is $\varepsilon/2$ in \cite{FRS,FRS2}.

Although the expression for $I_1 \sim I_{14}$ are given in \cite{FRS2},
some of them contain misprints.
Furthermore, 
when we calculate 
the anomalous dimension for the
tensor charge,
we need to prepare
some of $I_1\sim I_{14}$ separately 
as in the case of $I_{15,16}$.
In the following, we present such integral formulas.
\begin{eqnarray}
I_{3}(0,N)&=&{\hat{p}^{N}\over 2(16\pi^2)^2}\frac{1}{N+1}
\left[{1\over\varepsilon^2}
-{1\over\varepsilon}S_{1}(N+1)\right],\qquad\qquad(N\ge 0)\\[10pt]
I_{6}(M,0) &\equiv& \int\Dk
\frac{\konehat^{M}\ktwo\cdot p}{\kone^{4}(\kone-p)^{2}
      (\ktwo-p)^{2}(\kone-\ktwo)^{2}} \nonumber \\
      &=&  \frac{\phat^M}{4(16\pi^{2})^{2}}\frac{1}{M+1}
           \left[\;-{1\over 2\varepsilon^2}
           + {1\over\varepsilon}
   \left\{\frac{1}{2(M+1)}+\frac{1}{2M}-1\right\}\right],
\;\;\;(M\ge 1)\\[10pt]
I_{8}(M,N) &\equiv& \int\Dk
\frac{\konehat^{M}\ktwohat^{N}\kone\cdot p}{\kone^{4}\ktwo^{2}
(\ktwo-p)^{2}(\kone-\ktwo)^{2}}\nonumber \\
I_{8}(M,N)&=& \frac{\phat^{M+N}}{4(16\pi^{2})^{2}}\frac{1}{(M+1)(M+N)}
          \left[\;-{1\over \varepsilon^2}\right.
                +\left.{1\over\varepsilon}\right\{
            S_{1}(M+N)-S_{1}(M) \nonumber \\
      &&  +\left.\left.{1\over M+N+1}-{1\over M}
           -\frac{2}{M+1}\right\}\right],\qquad\qquad
(M\ge 1,\ N\ge 0)\\[10pt]
I_{8}(0,N)&=&{-\hat{p}^{N}\over 4(16\pi^2)^2}
           \frac{1}{N+1}{1\over\varepsilon},\qquad\qquad(N\ge 0)\\[10pt]
I_{11}(M,N) &\equiv& \int\Dk
\frac{\konehat^{M}\ktwohat^{N}}{\ktwo^{2}(\ktwo-p)^{2}
(\kone-\ktwo)^{2}(\kone -\ktwo +p)^{2}}\nonumber \\
      &=& \frac{-\phat^{M+N}}{(16\pi^{2})^{2}}{1\over\varepsilon^2}
          \frac{1}{M+N+2}
          \left[\;\frac{1}{M+1}+\frac{(-1)^{M}M!N!}
           {(M+N+1)!}\;\right].
\end{eqnarray}
The summation in $I_1$ and $I_{13}$ should start from $J=0$
in (A.2) and (A.14) of \cite{FRS2}.

All the other integral formula for $I_1\sim I_{14}$
which are necessary in our calculation
but not presented explicitly in the above
can be taken from Appendix of \cite{FRS2}.

\section{Anomalous dimension from each diagram}
\setcounter{equation}{0}
\renewcommand{\theequation}{B.\arabic{equation}}

In this appendix, we present the contribution
to $\gamma^{(1)}_n$ for $h_1$ from each Feynman diagram shown in Fig. 2,
for which we use the notation $\gamma^{(a)}_n$,
$\gamma^{(b)}_n$ {\it etc} with
obvious definition.
We also use the symbols $I_1^{D.P.}$ and 
$I_1^{S.P.}$ {\it etc} 
to designate, respectively, the coefficients of the 
$1/(16\pi^2)^2\varepsilon^2$ (double pole) term and the
$1/(16\pi^2)^2\varepsilon$ (simple pole) term of 
the integrals 
$I_1$ through $I_{16}$ defined in the appendix of \cite{FRS2}
and the appendix A above.

As noted in the text, $\gamma^{(a,b,c,p)}_n=0$, 
and $\gamma^{(d,g,h,l,m,q,r)}_n$ are the same as those for $f_1$
calculated in \cite{FRS}.
Other results are as follows: ($C_F=(N_c^2 -1)/2N_c$,
$C_G=N_c$, $T_R=N_f/2$ for the color group $SU(N_c)$ ):

\begin{eqnarray}
\gamma^{(e)}_n=-\frac{16}{9}C_{F}T_R.
\end{eqnarray}

\begin{eqnarray}
\gamma^{(f)}_n={32\over 9}C_{F}C_{G}.
\end{eqnarray}

{}Following the calculational procedure described in section 3,
we obtained very cumbersome results for 
$\gamma^{(i,j,k,n,o)}_n$ in terms of the integral formulas,
$I_1 \sim I_{16}$.  
It turned out, however, we can further simplify those results
as was done in \cite{GLY} for $f_1$.
In the following, we present the simplified forms together with
the expression with $I_1 \sim I_{16}$.
Although we had to 
calculate the integral formulas for $\gamma_{n=0}^{h(1)}$ 
separately from those for
$\gamma_{n\ge 1}^{h(1)}$, the final results for (i), (j), (k) 
can be written in a very simple form for all $n$, 
which supports the correctness of the calculation.

\begin{eqnarray}
\gamma^{(i)}_n
&=&16\left(C_{F}^{2}-\frac{1}{2}C_{F}C_{G}\right)
\phat^{-n}\biggl.\biggr[\phat^{-2}\biggl.\biggr\{-I_{1}^{S.P.}(n,2)
+2I_{1}^{S.P.}(1,n+1)+I_{2}^{S.P.}(n,2)
\nonumber\\
&&
-2I_{2}^{S.P.}(1,n+1)
-2I_{3}^{S.P.}(n+1,1)
+2I_{4}^{S.P.}(n+2,0)
-2I_{6}^{S.P.}(n+2,0)
\nonumber\\
&&
+2I_{8}^{S.P.}(n,2)\biggl.\biggr\}
+\phat^{-1}\biggl.\biggr\{I_{1}^{S.P.}(n,1)
-I_{1}^{S.P.}(0,n+1)
-I_{2}^{S.P.}(n,1)
+2I_{2}^{S.P.}(1,n)
\nonumber\\
&&
-I_{2}^{S.P.}(0,n+1)
-2I_{3}^{S.P.}(n,1)
+2I_{3}^{S.P.}(n+1,0)\biggl.\biggr\}
+2I_{1}^{S.P.}(0,n)
-2I_{15}^{S.P.}(n,0)
\nonumber\\
&&
+\phat^{-1}\biggl.\biggr\{I_{1}^{D.P.}(n,1)
-I_{1}^{D.P.}(0,n+1)
-2I_{1}^{D.P.}(1,n)
-I_{2}^{D.P.}(n,1)
-I_{2}^{D.P.}(0,n+1)
\nonumber\\
&&
+2I_{2}^{D.P.}(1,n)
+4I_{6}^{D.P.}(n+1,0)\biggl.\biggr\}
-I_{1}^{D.P.}(n,0)
+I_{1}^{D.P.}(0,n)
+I_{9}^{D.P.}(n,0)
\biggl.\biggr]\nonumber\\[10pt]
&=& -8\left(C_{F}^{2}-\frac{1}{2}C_{F}C_{G}\right)
\frac{1}{n+1}\biggl.\biggr(
S_{1}(n)+\frac{1}{n+1}\biggl.\biggr).
\end{eqnarray}

\begin{eqnarray}
\gamma^{(j)}_n
&=&4C_{F}C_{G}\phat^{-n}
\biggl.\biggr[2\phat^{-2}
\biggl.\biggr\{I_{1}^{S.P.}(n,2)
-2I_{1}^{S.P.}(1,n+1)
-I_{2}^{S.P.}(n,2)
+2I_{2}^{S.P.}(1,n+1)
\nonumber\\
&&
+2I_{3}^{S.P.}(n+1,1)
-2I_{4}^{S.P.}(n+2,0)
+2I_{6}^{S.P.}(n+2,0)
-2I_{8}^{S.P.}(n,2)\biggl.\biggr\}
\nonumber\\
&&
+\phat^{-1}\biggl.\biggr\{
-3I_{1}^{S.P.}(n,1)
+3I_{1}^{S.P.}(0,n+1)
+3I_{2}^{S.P.}(n,1)
-4I_{2}^{S.P.}(1,n)
+I_{2}^{S.P.}(0,n+1)
\nonumber\\
&&
+4I_{3}^{S.P.}(n,1)
-4I_{3}^{S.P.}(n+1,0)\biggl.\biggr\}
+I_{1}^{S.P.}(n,0)
-3I_{1}^{S.P.}(0,n)
+4I_{15}^{S.P.}(n,0)
\nonumber\\
&&
+\phat^{-1}\biggl.\biggr\{
-I_{1}^{D.P.}(n,1)
+I_{1}^{D.P.}(0,n+1)
+4I_{1}^{D.P.}(1,n)
+I_{2}^{D.P.}(n,1)
+3I_{2}^{D.P.}(0,n+1)
\nonumber\\
&&
-4I_{2}^{D.P.}(1,n)
-8I_{6}^{D.P.}(n+1,0)\biggl.\biggr\}
+I_{1}^{D.P.}(n,0)
-3I_{1}^{D.P.}(0,n)
-I_{9}^{D.P.}(n,0)
\biggl.\biggr]\nonumber\\[5pt]
&=& 0.
\end{eqnarray}
The contribution from (j)
turned out to be identically zero for all $n$.

\begin{eqnarray}
\gamma^{(k)}_n
&=&  8\left(C_{F}^{2}-\frac{1}{2}C_{F}C_{G}\right)
\phat^{-n}\biggl.\biggr[\;\phat^{-2}(1+(-1)^{n})
\biggl.\biggr\{
I_{3}^{S.P.}(n+2,0)
-2I_{8}^{S.P.}(n,2)
-2I_{16}^{S.P.}(n,2)
\nonumber\\
&&
+4I_{16}^{S.P.}(n+1,1)
-2I_{16}^{S.P.}(n+2,0)
\biggl.\biggr\}
+\phat^{-1}\biggl.\biggr\{
I_{2}^{S.P.}(n,1)
+I_{12}^{S.P.}(n,1)
-I_{12}^{S.P.}(n+1,0)
\nonumber\\
&&
-I_{13}^{S.P.}(n,1)
+I_{13}^{S.P.}(n+1,0)
-I_{14}^{S.P.}(n,1)
+4(1+(-1)^{n})I_{16}^{S.P.}(n,1)
\nonumber\\
&&
-4(1+(-1)^{n})I_{16}^{S.P.}(n+1,0)\biggl.\biggr\}
+(1+(-1)^{n})I_{3}^{S.P.}(n,0)
-(-1)^{n}I_{12}^{S.P.}(n,0)
+2I_{13}^{S.P.}(n,0)
\nonumber\\
&&
+I_{14}^{S.P.}(n,0)
-2(1+(-1)^{n})I_{16}^{S.P.}(n,0)
+\phat^{-1}\biggl.\biggr\{
I_{2}^{D.P.}(n,1)
+I_{12}^{D.P.}(n,1)
-I_{12}^{D.P.}(n+1,0)
\nonumber\\
&&
-I_{13}^{D.P.}(n,1)
+I_{13}^{D.P.}(n+1,0)
-I_{14}^{D.P.}(n,1)\biggl.\biggr\}
-I_{11}^{D.P.}(n,0)
+I_{13}^{D.P.}(n,0)
+I_{14}^{D.P.}(n,0)
\biggl.\biggr]\nonumber\\[10pt]
&=&
-8\left(C_{F}^{2}-\frac{1}{2}C_{F}C_{G}\right)
(1+(-1)^{n})
\biggl.\biggr(
\frac{1}{n+1} - \frac{1}{n+2}\biggl.\biggr).
\end{eqnarray}
This result is zero for odd-$n$.  This can be easily proved
by assigning the momenta as shown in Fig. 10.
Then the denominator is symmetric under $k_1 \leftrightarrow
k_2$.  The numerator has a factor $(\hat{k}_1-\hat{k}_2)^n$.  
It is convenient to consider
the sum of Fig. 10 and its charge conjugation.  Then one can
easily show the contribution from (k) becomes 0 for odd-$n$
with the use of lemma in section 3.

\begin{eqnarray}
\gamma^{(n)}_n
&=&
8C_{F}C_{G}\phat^{-n}
\sum_{l=1}^{n}\biggl.\biggr[\;\phat^{-1}\biggl.\biggr\{
-I_{1}^{S.P.}(l-1,n-l+2)
+I_{1}^{S.P.}(l,n-l+1)
\nonumber\\
&&
+I_{2}^{S.P.}(l-1,n-l+2)
-I_{2}^{S.P.}(l,n-l+1)\biggl.\biggr\}
+I_{1}^{S.P.}(l,n-l)
\biggl.\biggr]\nonumber\\[10pt]
&=& 4C_FC_G \left[S_3(n)-2\widehat{S}(n) -{S_1^2(n)\over 2(n+1)}
-{S_2(n)\over 2(n+1)} + {(2n-1)S_1(n)\over (n+1)^2}
+{2n\over(n+1)^3}\right].\nonumber\\
\end{eqnarray}

\begin{eqnarray}
\gamma^{(o)}_n &=&
-16\left(C_{F}^{2}-\frac{1}{2}C_{F}C_{G}\right)
\phat^{-n}\sum_{l=1}^{n}
\biggl.\biggr[\;
\phat^{-1}\biggl.\biggr\{
-I_{2}^{S.P.}(n-l,l+1)
+I_{2}^{S.P.}(n-l+1,l)
\nonumber\\
&&
+I_{12}^{S.P.}(n-l,l+1)
-2I_{12}^{S.P.}(n-l+1,l)
+I_{12}^{S.P.}(n-l+2,l-1)
\nonumber\\
&&
-I_{13}^{S.P.}(n-l,l+1)
+2I_{13}^{S.P.}(n-l+1,l)
-I_{13}^{S.P.}(n-l+2,l-1)
\nonumber\\
&&
+I_{14}^{S.P.}(n-l,l+1)
-I_{14}^{S.P.}(n-l+1,l)\biggl.\biggr\}
+I_{13}^{S.P.}(n-l,l)
-I_{13}^{S.P.}(n-l+1,l-1)
\nonumber\\
&&
-I_{14}^{S.P.}(n-l,l)
-I_{14}^{S.P.}(n-l+1,l-1)
\biggl.\biggr]\nonumber\\[10pt]
&=& 
-8\left( C_F^2- {1\over 2}C_FC_G\right)
\left[
4\left( S_1(n) + {1\over n+1}
\right)\left(S_2'\left({n\over 2}\right)
-S_2(n)\right)
- 4 \widehat{S}(n) -8\widetilde{S}(n)\right.\nonumber\\[10pt]
& &\left.+ 2 S_3(n) +
S_3'\left({n\over 2}\right)
-{ 2 S_1(n)\over n+1} + 
{4 \left( 1-(-1)^n\right)
\over (n+1)^3 }-{2n\over (n+1)^3 }\right].
\end{eqnarray}
In the above expression, we introduced the function
\beq
\widehat{S}(n) = \sum_{j=1}^n{1\over j^2}S_1(j).
\eeq

\newpage

\Large
\centerline{\bf Figure captions}

\normalsize
\vspace{0.5cm}

\begin{enumerate}

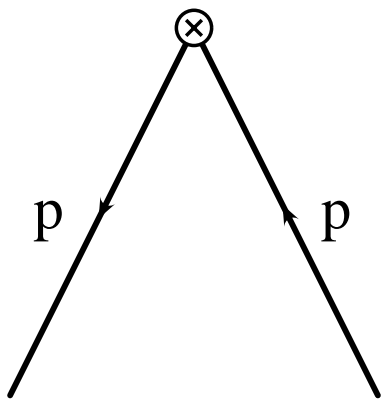
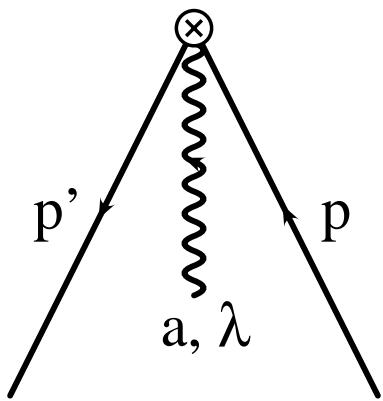
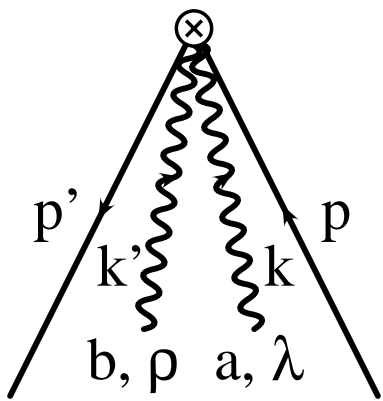
\item[{\bf Fig. 1}]
Basic vertices for 
$O_n^\mu\cdot\Delta$ with
$0$, $1$ and $2$ gluon lines.

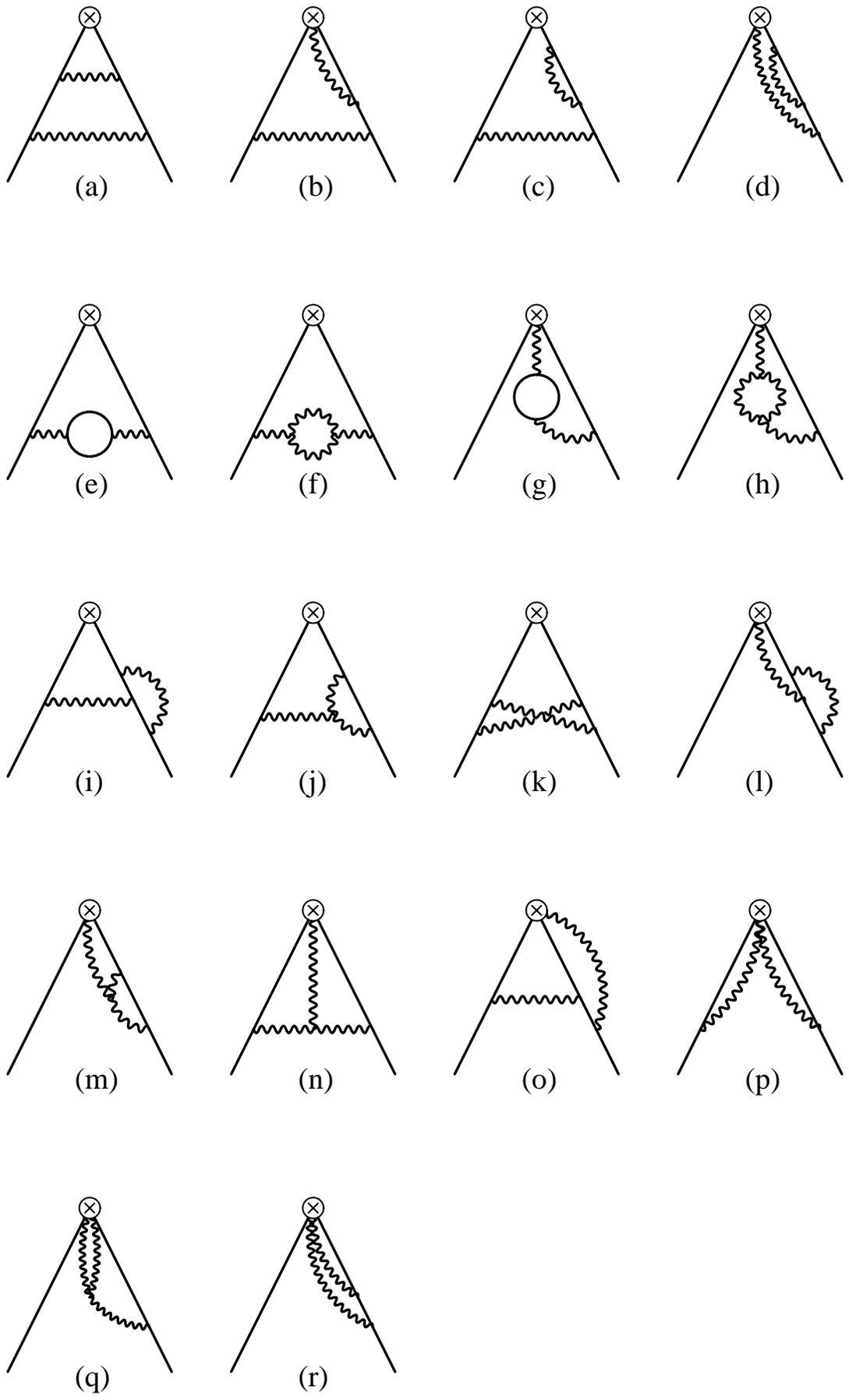
\item[{\bf Fig. 2}]
Two-loop corrections to
the one-particle irreducible
two-point Green function which imbeds the operator 
$O_n^\mu\cdot\Delta$.
For diagrams not symmetric, crossed diagrams must be included.
In Figs. (f) and (h), the contributions from the Faddev-Popov ghost and
tadpole diagram must be added.

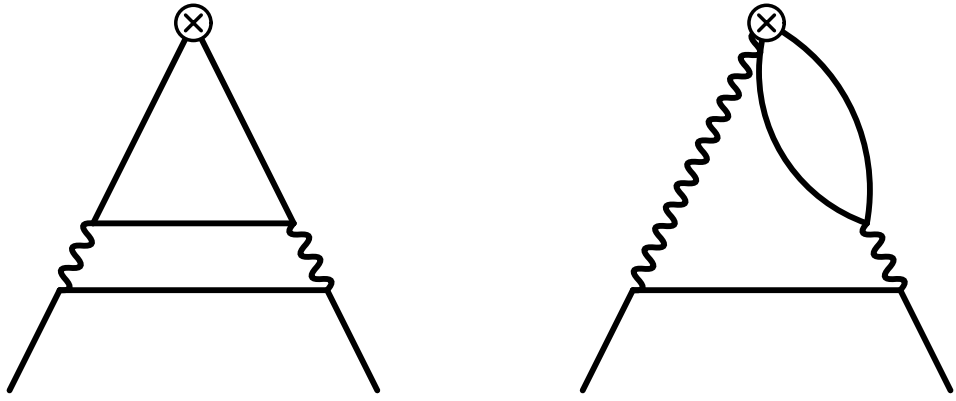
\item[{\bf Fig. 3}]
The same as Fig. 2 but for the flavor-singlet distribution.
These diagrams become identically zero for $h_1$.

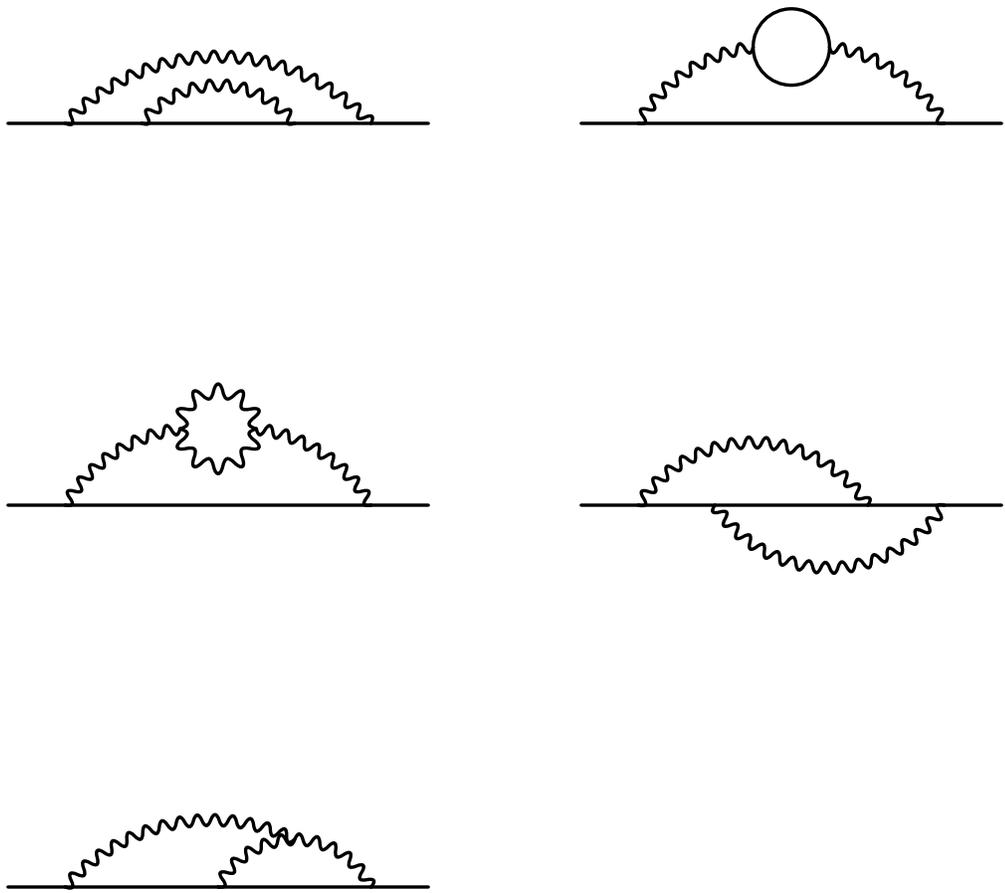
\item[{\bf Fig. 4}]
Two-loop corrections to the quark self-energy.  The ghost-loop
and the tadpole contribution is implied as in Fig. 2.

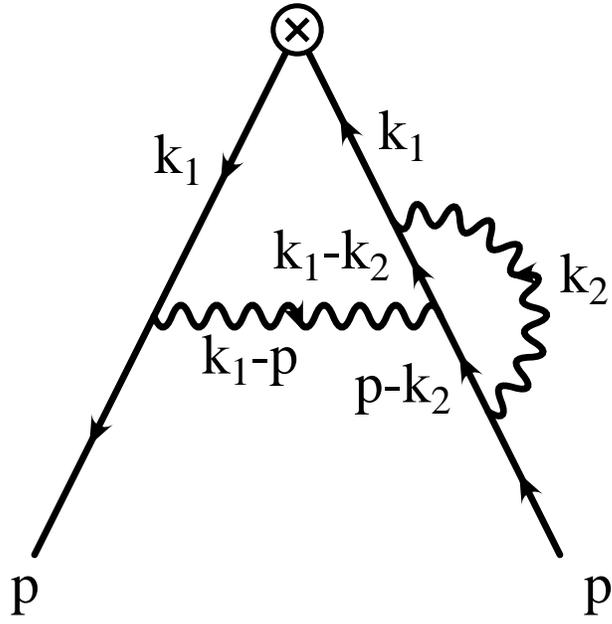
\item[{\bf Fig. 5}]
Momentum labeling for the diagram Fig. 2(i).

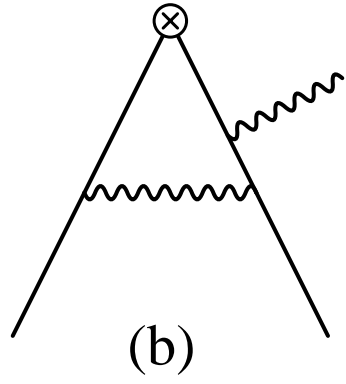
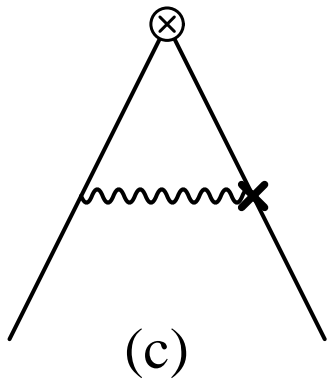
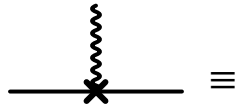
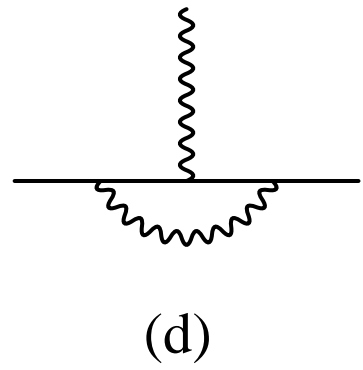
\item[{\bf Fig. 6}]
The counter term contributions to Fig. 2(i).

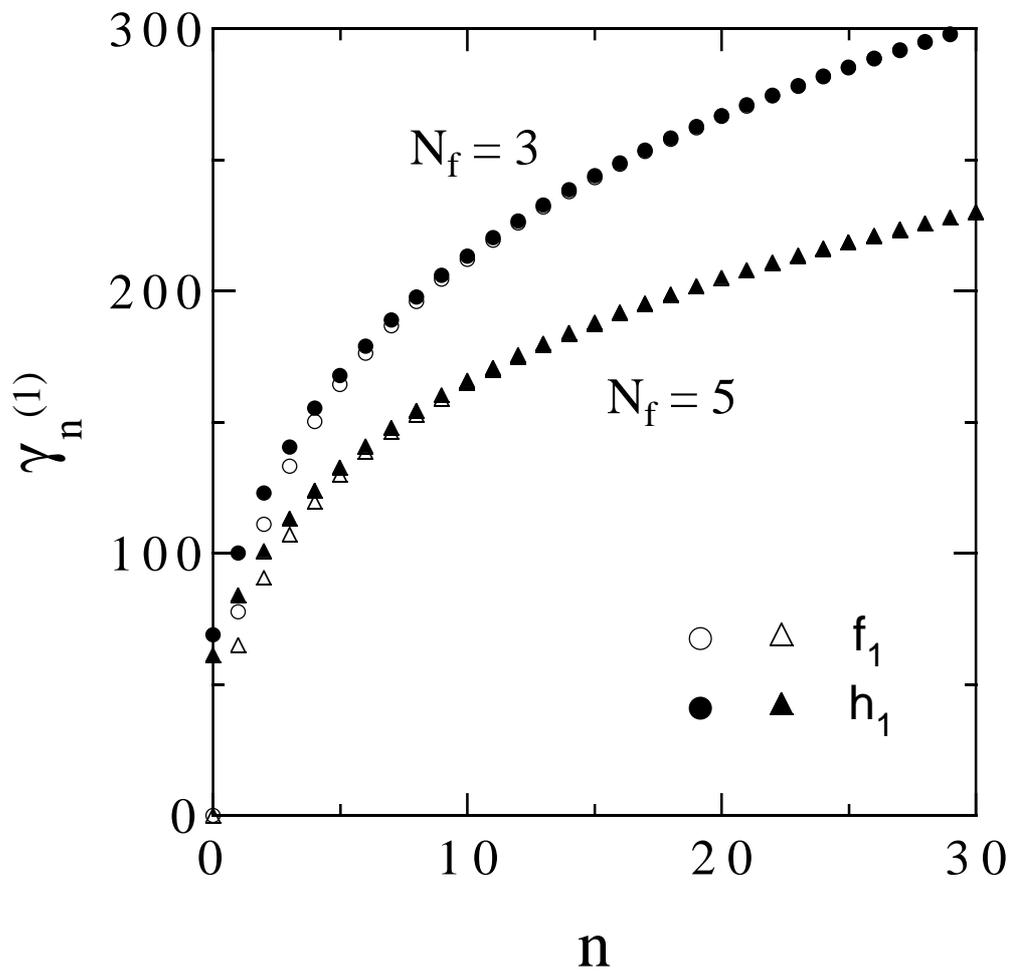
\item[{\bf Fig. 7}]
The coefficients of the two-loop 
anomalous dimension $\gamma^{f,h(1)}_n$ for $N_f=3$ (circle)
and $N_f=5$ (triangle).

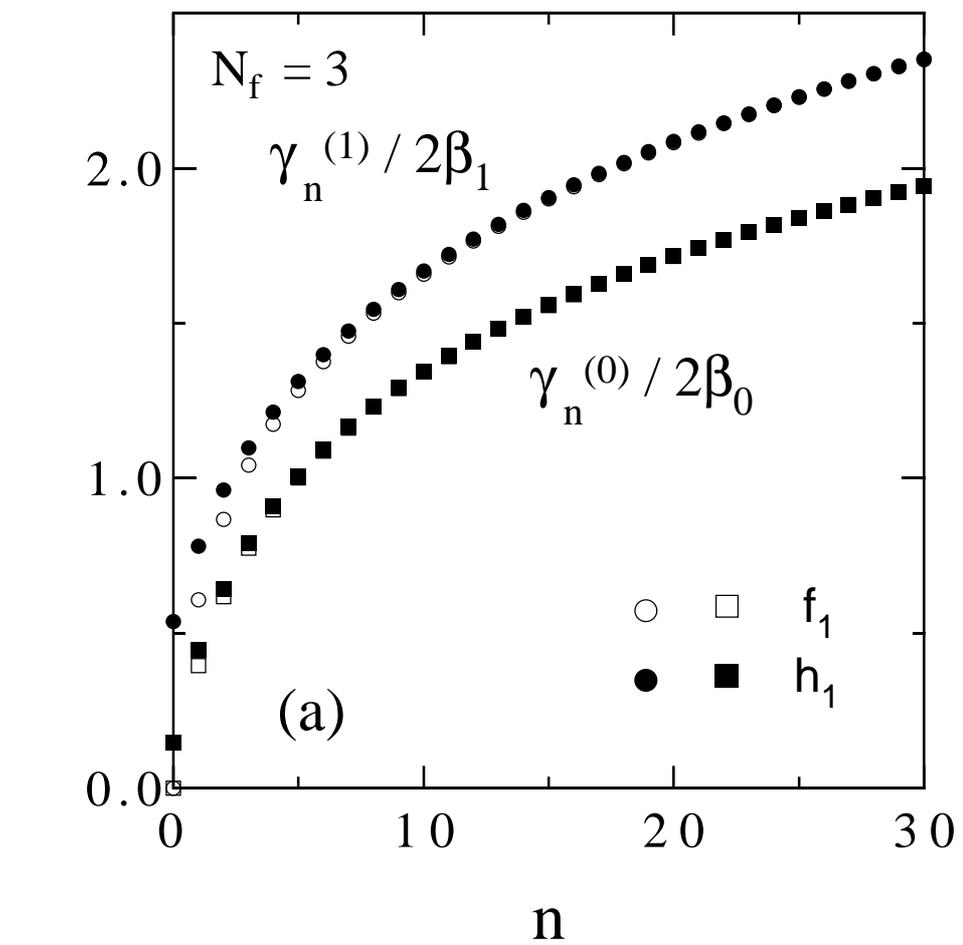
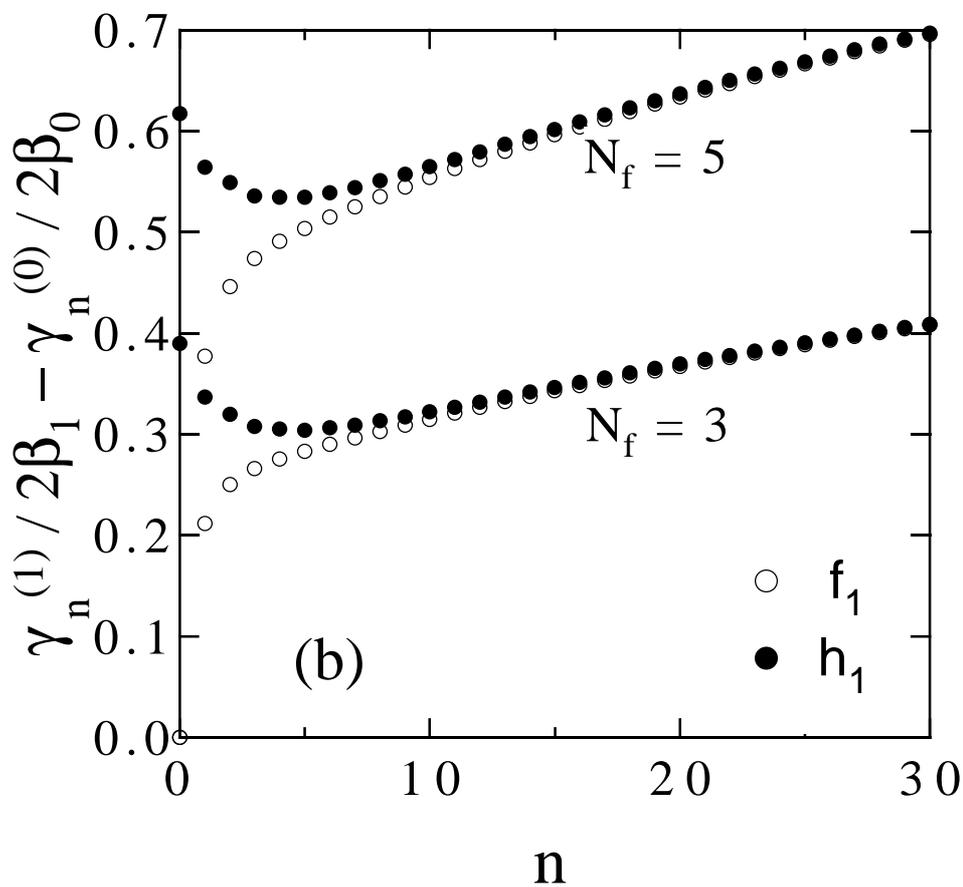
\item[{\bf Fig. 8}]
(a) $\gamma^{f,h(1)}_n/2\beta_1$ (circle) and
$\gamma^{f,h(0)}_n/2\beta_0$ (square).
(b) $\gamma^{f,h(1)}_n/2\beta_1 - \gamma^{f,h(0)}_n/2\beta_0$ for
$N_f=3,\ 5$.

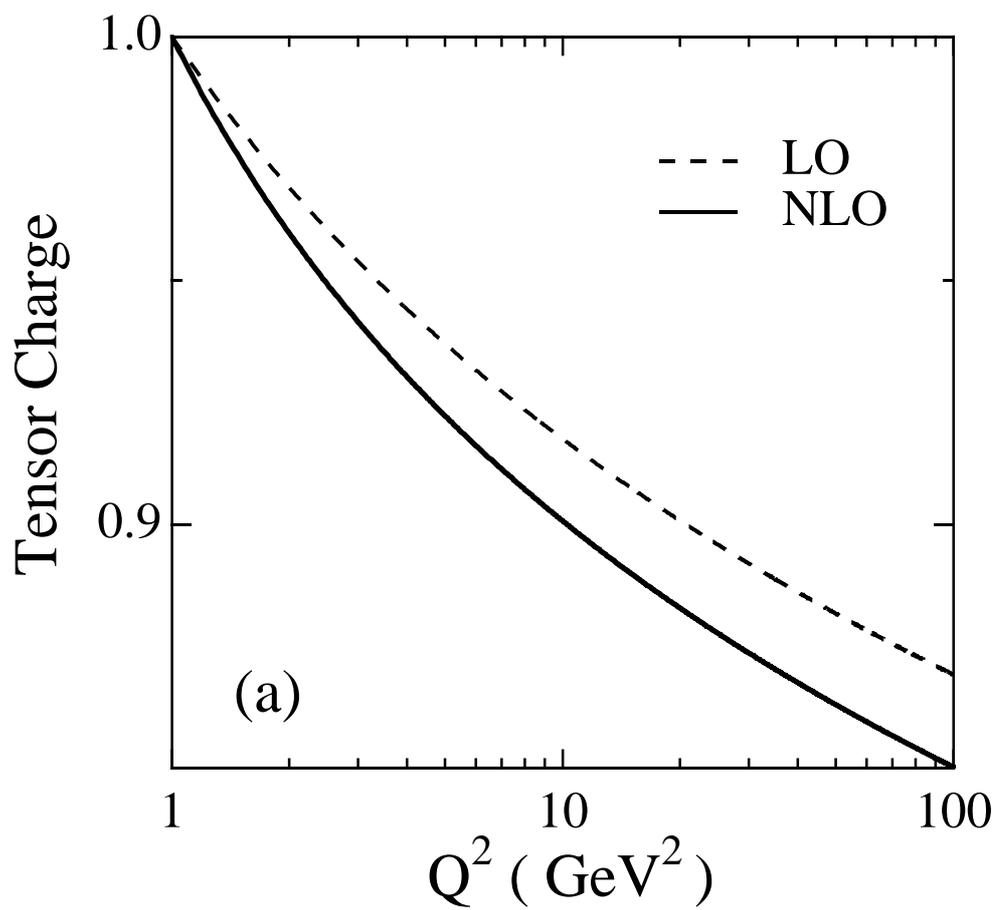
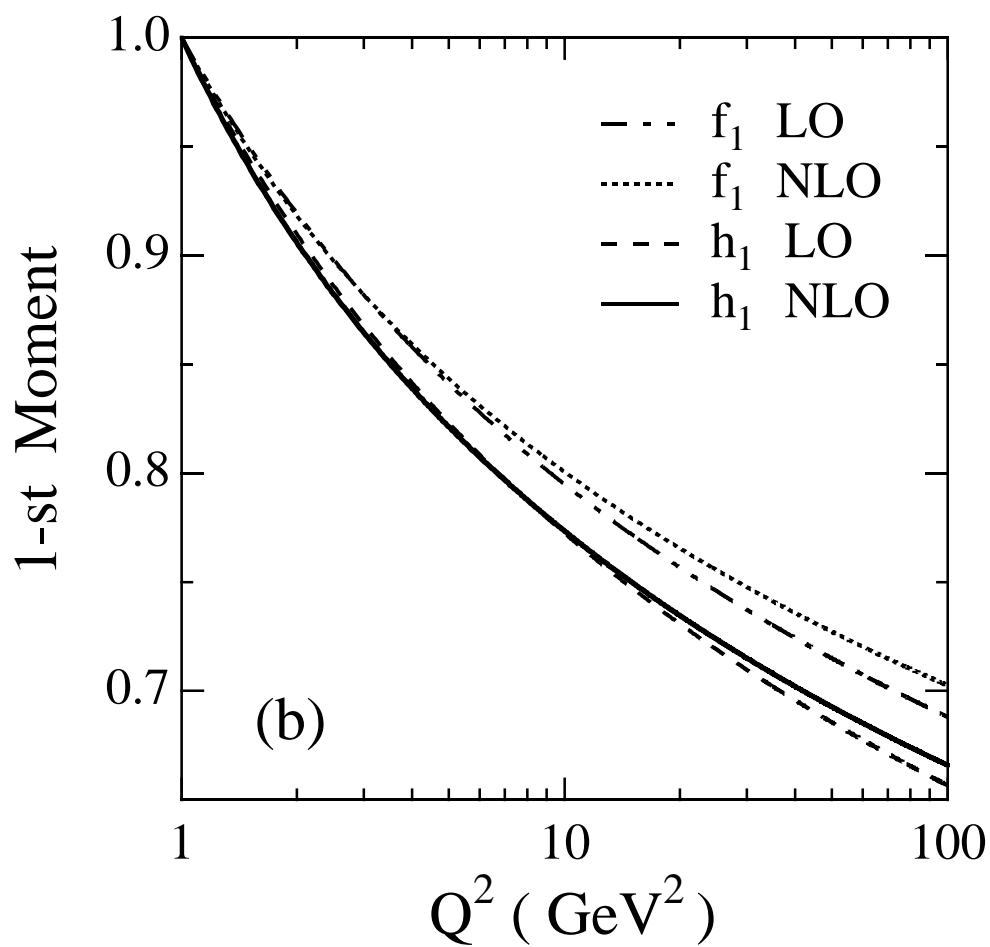
\item[{\bf Fig. 9}]
(a) The LO and NLO $Q^2$-evolution of the tensor charge
normalized at $Q^2=1$ GeV$^2$.
(b) The LO and NLO $Q^2$-evolution of the first moment of 
$h_1(x,Q^2)$ and $f_1(x,Q^2)$ normalized at $Q^2=1$ GeV$^2$.

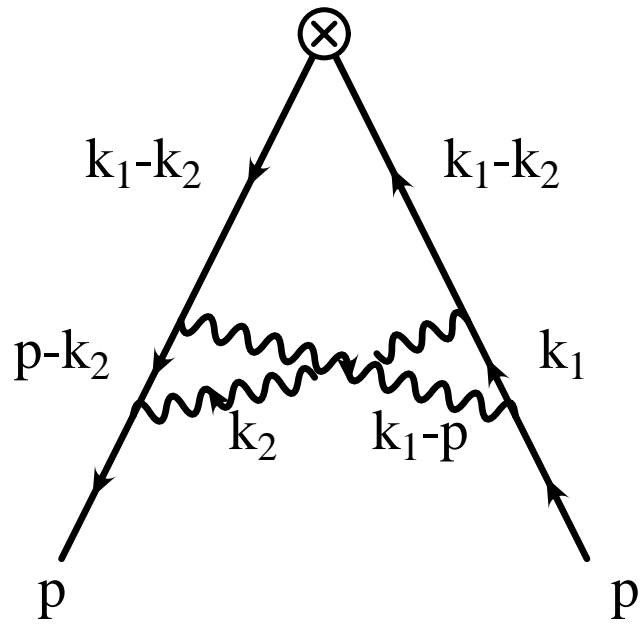
\item[{\bf Fig. 10}]
Momentum labeling for the diagram Fig. 2(k).

\end{enumerate}

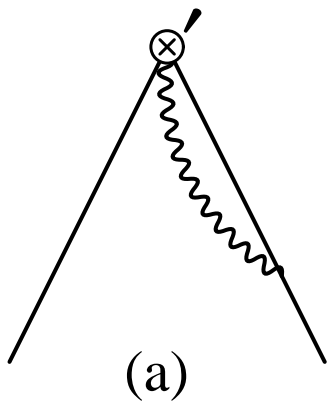
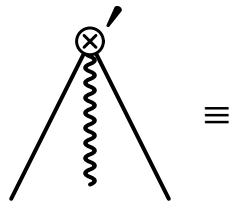
\end{document}